\documentclass[aps,showpacs,preprintnumbers,amsmath, amssymb]{revtex4}

\usepackage{float}
\usepackage{graphics,epsfig}
\usepackage{graphicx}
\usepackage{dcolumn}
\usepackage{bm}

\begin{document}
\baselineskip=0.8 cm
\title{{\bf Wave dynamics of phantom scalar perturbation
in the background of Schwarzschild black hole}}
\author{Songbai Chen}
\email{csb3752@163.com} \affiliation{Institute of Physics and
Department of Physics,
Hunan Normal University,  Changsha, Hunan 410081, P. R. China \\
Key Laboratory of Low Dimensional Quantum Structures and Quantum
Control (Hunan Normal University), Ministry of Education, P. R.
China.}

\author{Jiliang Jing }
\email{jljing@hunnu.edu.cn}
 \affiliation{Institute of Physics and
Department of Physics,
Hunan Normal University,  Changsha, Hunan 410081, P. R. China \\
Key Laboratory of Low Dimensional Quantum Structures and Quantum
Control (Hunan Normal University), Ministry of Education, P. R.
China.}

\author{Qiyuan Pan}
\email{panqiyuan@126.com}
 \affiliation{Institute of Physics and
Department of Physics,
Hunan Normal University,  Changsha, Hunan 410081, P. R. China \\
Key Laboratory of Low Dimensional Quantum Structures and Quantum
Control (Hunan Normal University), Ministry of Education, P. R.
China.}

\vspace*{0.2cm}
\begin{abstract}
\baselineskip=0.6 cm
\begin{center}
{\bf Abstract}
\end{center}

Using Leaver's continue fraction and time domain method, we
investigate the wave dynamics of phantom scalar perturbation in the
background of Schwarzschild black hole.  We find that the presence
of the negative kinetic energy terms modifies the standard results
in quasinormal spectrums and late-time behaviors of the scalar
perturbations. The phantom scalar perturbation in the late-time
evolution will grow with an exponential rate.
\end{abstract}

\maketitle
\newpage
\vspace*{0.2cm}

Many observations confirm that our universe is undergoing an
accelerated expansion. In order to explain this observed phenomena,
the universe is supposed to be filled with dark energy within the
framework of Einstein's general relativity. Dark energy is an exotic
energy component with negative pressure and constitutes about $72\%$
of present total cosmic energy. The simplest explanation for dark
energy is the cosmological constant \cite{1a}, which is a term that
can be added to Einstein's equations. This term acts like a perfect
fluid with an equation of state $\omega_x=-1$, and the energy
density is associated with quantum vacuum. Although this
interpretation is consistent with observational data, it suffers the
coincidence problem, namely, ``why are the vacuum and matter energy
densities of precisely the same order today?". Therefore the
dynamical scalar fields, such as quintessence \cite{2a}, k-essence
\cite{3a} and phantom field \cite{4a}, have been put forth as an
alternative of dark energy.

The phantom field model is an interesting candidate for dark energy
since it has some peculiar properties. This field has a negative
kinetic energy and so that the null energy condition is violated and
the equation of state $\omega_x$ of dark energy less than $-1$. The
super negative equation of state is favored by recent precise
observational data involving CMB, Hubble Space Telescope, type Ia
Supernova, and 2dF data sets \cite{5a}. The dynamical properties of
the phantom field in the cosmology has been investigated in the last
years \cite{6a1,6a11,6a2,6a3,6a4,6a5,6a6,6a7}. It shows that the
energy density increases with the time and approaches to infinity in
a finite time \cite{6a1}. In other words, the universe dominated by
phantom energy will blow up incessantly and arrive at a big rip
finally, which is a future singularity with a strong exclusive force
so that anything in the universe including the large galaxies will
be torn up. Recently, many efforts have been working to avoid the
big rip \cite{7b}. It has argued that if one considers the effects
from loop quantum gravity, this future singularity will be
disappeared in the universe \cite{lc3,lg4,lg6,lg7}.

It is of interest to extend the study of dynamical evolution of the
phantom field to black hole spacetime in various gravity theories,
since this could help us to obtain the connection between dark
energy and black hole. The dynamical evolution of usual scalar field
perturbation has been studied for the last few decades (for a
review, see \cite{qn1,qn2,qn3}). It is well known that a static
observer outside a black hole can indicate three successive stages
of the wave evolution. The first one is that the exact shape of the
wave front depends on the initial pulse. This stage is followed by a
quasinormal ringing, which describes the damped oscillations under
perturbations in the surrounding geometry of a black hole with
frequencies and damping times of the oscillations entirely fixed by
the black hole parameters. It is widely believed that the
quasinormal modes carry characteristic fingerprint of a black hole
and can offer a direct way to identify the black hole existence.
Detection of these quasinormal modes is expected to be realized
through gravitational wave observation in the near future
\cite{qn1,qn2}. Apart from the potential astrophysical interest,
quasinormal modes could also serve as a tool to test ground of
fundamental physics. It has been argued that the study of QNM can
help us get deeper understandings of the AdS/CFT \cite{qn3,qn4},
dS/CFT \cite{qn5} correspondences, loop quantum gravity \cite{qn6}
and also the phase transition of black holes \cite{qn7}. In this
letter, we treat the phantom field as an external perturbation and
examine whether there exists some new feature in the dynamical
evolution of phantom scalar field in a black hole spacetime.

In the curve spacetime, the action of the phantom scalar field with
the negative kinetic energy term is
\begin{eqnarray}
S=\int d^4x \sqrt{-g}\bigg[-\frac{R}{16\pi
G}-\frac{1}{2}\partial_{\mu}\psi\partial^{\mu}\psi-V(\psi)\bigg].
\end{eqnarray}
Here we take metric signature $(+ - - -)$. The usual ``Mexican hat"
symmetry breaking potential has the form
\begin{eqnarray}
V(\psi)=-\frac{1}{2}\mu^2\psi^2+\frac{\kappa}{4}\psi^4,
\end{eqnarray}
where $\mu$ is the mass of the scalar field and $\kappa$ is the
coupling constant. In general, the presence of the phantom scalar
field will change the structure of the black hole spacetime
\cite{phso1}. Here we just treat it as an external perturbation and
suppose it does not affect the metric of the background. Meanwhile
we only consider the case $\kappa=0$ for conveniently, namely, the
potential has the form $V(\psi)=-\frac{1}{2}\mu^2\psi^2$. The wave
dynamics of usual scalar perturbations with this type of the
potential form  has been extensively investigated in the various
black holes spacetime \cite{ml1,ml2}.

Varying the action with respect to $\psi$, we obtain the wave
equation for phantom scalar field in the curve spacetime
\begin{eqnarray}
\frac{1}{\sqrt{-g}}\partial_{\mu}(\sqrt{-g}g^{\mu\nu}\partial_{\nu})
\psi-\mu^2\psi=0.\label{WE}
\end{eqnarray}
Comparing the Klein-Gordon equation of usual massive scalar field,
we find the unique difference in equation (\ref{WE}) is that the
sign of the mass term $\mu^2$ is negative, which will yield the
peculiar dynamical evolution of the phantom perturbation in the
black hole spacetime.

Let us now to consider the case of a Schwarzschild black hole
spacetime, whose metric in the standard coordinate can be described
by
\begin{eqnarray}
ds^2=(1-\frac{2M}{r})dt^2-(1-\frac{2M}{r})^{-1}dr^2-r^2(d\theta^2+\sin^2\theta
d\phi^2).
\end{eqnarray}

Separating $\psi=e^{-i\omega t}R(r)Y_{lm}(\theta,\phi)/r$, we can
obtain the radial equation for the scalar perturbation in the
Schwarzschild black hole spacetime
\begin{eqnarray}
\frac{d^2 R(r)}{dr^2_*}+[\omega^2-V(r)]R(r)=0,\label{jw}
\end{eqnarray}
where $r_*$ is the tortoise coordinate (which is defined by
$dr_*=\frac{r}{r-2M}dr$) and the effective potential $V(r)$ reads
\begin{eqnarray}
V(r)=\bigg(1-\frac{2M}{r}\bigg)\bigg(\frac{l(l+1)}{r^2}+\frac{2M}{r^3}-\mu^2\bigg).\label{efp}
\end{eqnarray}
Obviously, as $\mu=0$ the radial equation of phantom scalar field
reduce to that of usual massless scalar field. In the case
$\mu\neq0$, the effective potential $V(r)$ vanishes at the event
horizon and approaches to a negative constant $-\mu^2$ at the
spatial infinity. This is different from that of usual massive
scalar perturbation. Moreover, from figure (1), one can see that as
the mass $\mu$ increases the effective potential $V(r)$ for the
phantom scalar perturbation decreases and for the usual one
increases. This implies the wave dynamics of the phantom scalar
perturbation possesses some different properties from that of the
usual scalar perturbation.
\begin{figure}[ht]
\begin{center}
\includegraphics[width=6.2cm]{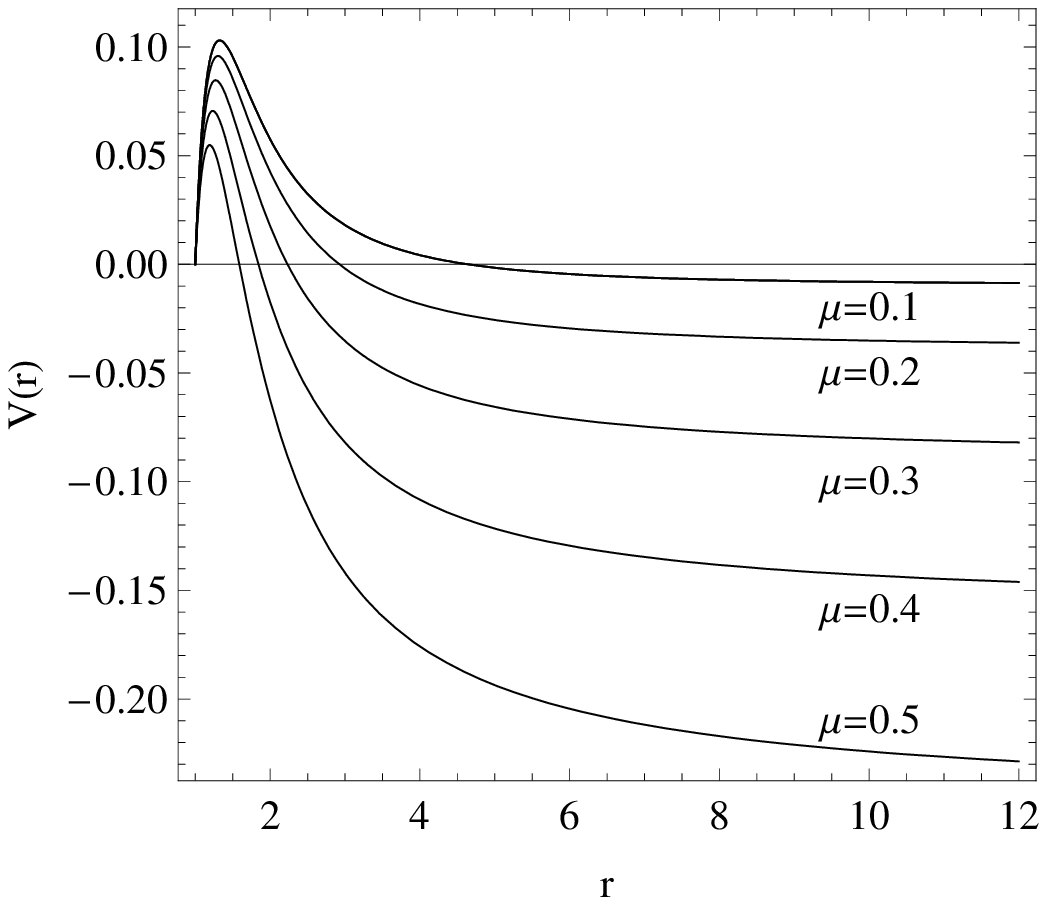}\ \ \ \ \ \ \includegraphics[width=6cm]{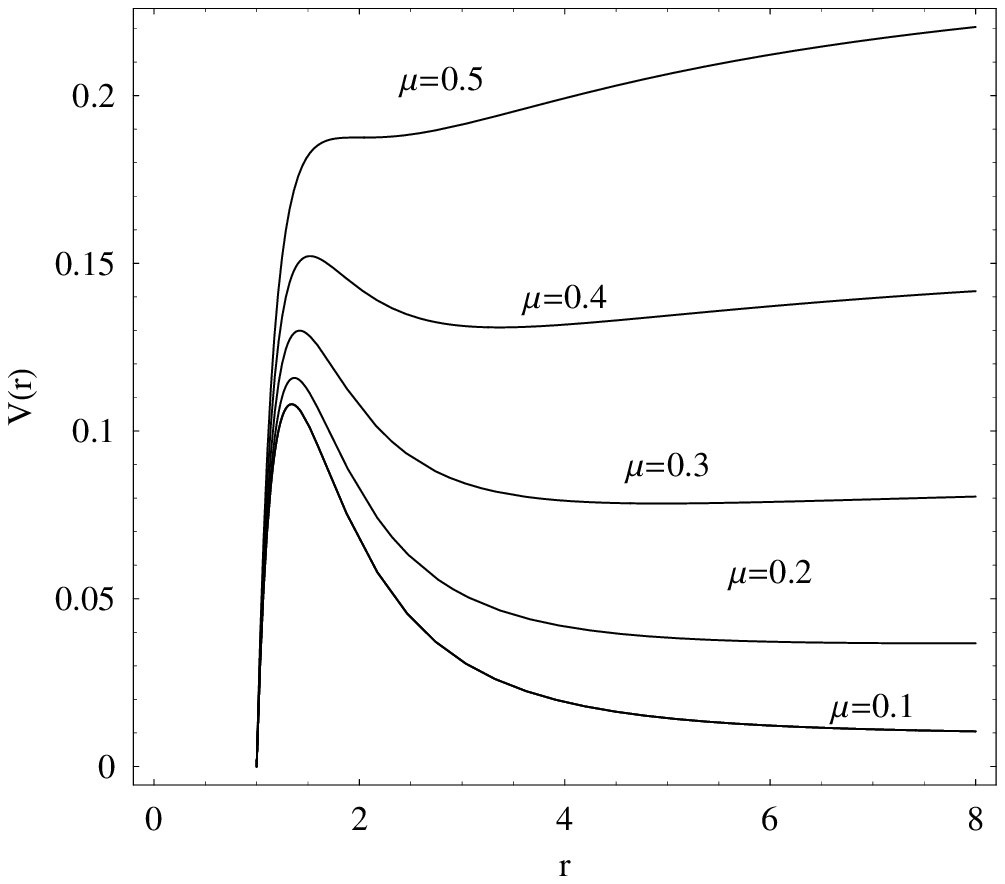}
\caption{Variety of the effective potential $V(r)$ with $r$ for
different $\mu$. The left for phantom scalar field and the right for
normal scalar field. Here $l=0$ and $M=0.5$.}
\end{center}
\label{fig1}
\end{figure}

We are now in a position to apply the continue fraction method
\cite{Leaver} and calculate the fundamental quasinormal modes
($n=0$) of phantom scalar perturbation in the Schwarzschild black
hole. From equation (\ref{efp}), we know that the boundary
conditions on the wave function $R(r)$ have the form
\begin{eqnarray}
R(r)=\bigg\{\begin{array}{lll}
(r-1)^{-i\omega},\;\;\;\;\;\;\;\;\;\;\;\;r\rightarrow 1,\\
\\
r^{\frac{i(\chi^2+\omega^2)}{2\chi}}e^{i\chi
r},\;\;\;\;\;\;\;\;\;r\rightarrow \infty,
\end{array}\label{bd}
\end{eqnarray}
where we set $2M=1$ and $\chi=\sqrt{\omega^2+\mu^2}$. A solution to
Eq.(\ref{jw}) that has the desired behavior at the boundary can be
written as
\begin{eqnarray}
R(r)=r^{i(\chi+\omega)-\frac{i\mu^2}{2\chi}}(r-1)^{-i\omega}e^{i\chi
r}\sum_{n=0}^{\infty}a_n\bigg(\frac{r-1}{r}\bigg)^n.\label{wbes}
\end{eqnarray}
The sequence of the expansion coefficients $\{a_n: n=1,\;2,
\;....\}$ is determined by a three-term recurrence relation staring
with $a_0=1$:
\begin{eqnarray}
&& \alpha_0 a_1+\beta_0a_0=0,\nonumber\\
&&
\alpha_na_{n+1}+\beta_na_n+\gamma_na_{n-1}=0,\;\;\;\;\;\;n=1,\;2,\;
\ldots.
\end{eqnarray}
The recurrence coefficient $\alpha_n$, $\beta_n$ and $\gamma_n$ are
given in terms of $n$ and the black hole parameters by
\begin{eqnarray}
&& \alpha_n=n^2+(C_0+1)n+C_0,\nonumber\\
&& \beta_n=-2n^2+(C_1+2)n+C_3,\nonumber\\
&& \gamma_n=n^2+(C_2-3)n+C_4-C_2+2,
\end{eqnarray}
and the intermediate constants $C_n$ are defined by
\begin{eqnarray}
&& C_0=1-2i\omega,\nonumber\\
&& C_1=-4+i(4\omega+3\chi)+\frac{i\omega^2}{\chi},\nonumber\\
&& C_2=3-i(2\omega+\chi)-\frac{i\omega^2}{\chi},\nonumber\\
&& C_3=\bigg(\frac{\omega+\chi}{\chi}\bigg)\bigg[(\omega+\chi)^2+\frac{i(\omega+3\chi)}{2}\bigg]-l(l+1)-1,\nonumber\\
&& C_4=-\bigg[\frac{(\omega+\chi)^2}{2\chi}+i\bigg]^2.
\end{eqnarray}
If the boundary condition (\ref{bd}) is satisfied and the series in
(\ref{wbes}) converge for the given $l$, the frequency $\omega$ is a
root of the continued fraction equation
\begin{eqnarray}
\beta_n-\frac{\alpha_{n-1}\gamma_n}{\beta_{n-1}-\frac{\alpha_{n-2}\gamma_{n-1}}{\beta_{n-2}
-\alpha_{n-3}\gamma_{n-2}/...}}=\frac{\alpha_{n}\gamma_{n+1}}{\beta_{n+1}-\frac{\alpha_{n+1}\gamma_{n+2}}{\beta_{n+2}
-\alpha_{n+2}\gamma_{n+3}/...}}.
\end{eqnarray}
This equation is impossible to be solved analytically. But we can
rely on the numerical calculation to obtain the quasinormal
frequencies for phantom scalar perturbations in the Schwarzschild
black hole spacetime.
\begin{table}[h]
\begin{center}
\begin{tabular}[b]{cccc}
 \hline \hline
 \;\;\;\; $\mu$ \;\;\;\; & \;\;\;\; $l=0$\;\;\;\;  & \;\;\;\;  $l=1$\;\;\;\;
 & \;\;\;\; $l=2$ \;\;\;\; \\ \hline
\\
0& \;\;\;\;\;0.220910-0.209791i\;\;\;\;\;  & \;\;\;\;
0.585872-0.19532i\;\;\;\;\;
 & \;\;\;\;\;0.967288-0.193518i\;\;\;\;\;
 \\
0.1&0.223315-0.227501i&0.579967-0.200625i&0.962955-0.195599i
 \\
0.2&0.228479-0.269382i&0.562925-0.217049i&0.949962-0.201995i
 \\
0.3&0.232013-0.321298i&0.537016-0.245696i&0.928354-0.213175i
\\
\hline \hline
\end{tabular}
\end{center}
\caption{The fundamental ($n=0$) quasinormal frequencies of phantom
scalar field in the Schwarzschild black hole spacetime.}
\end{table}
\begin{figure}[ht]
\begin{center}
\includegraphics[width=6cm]{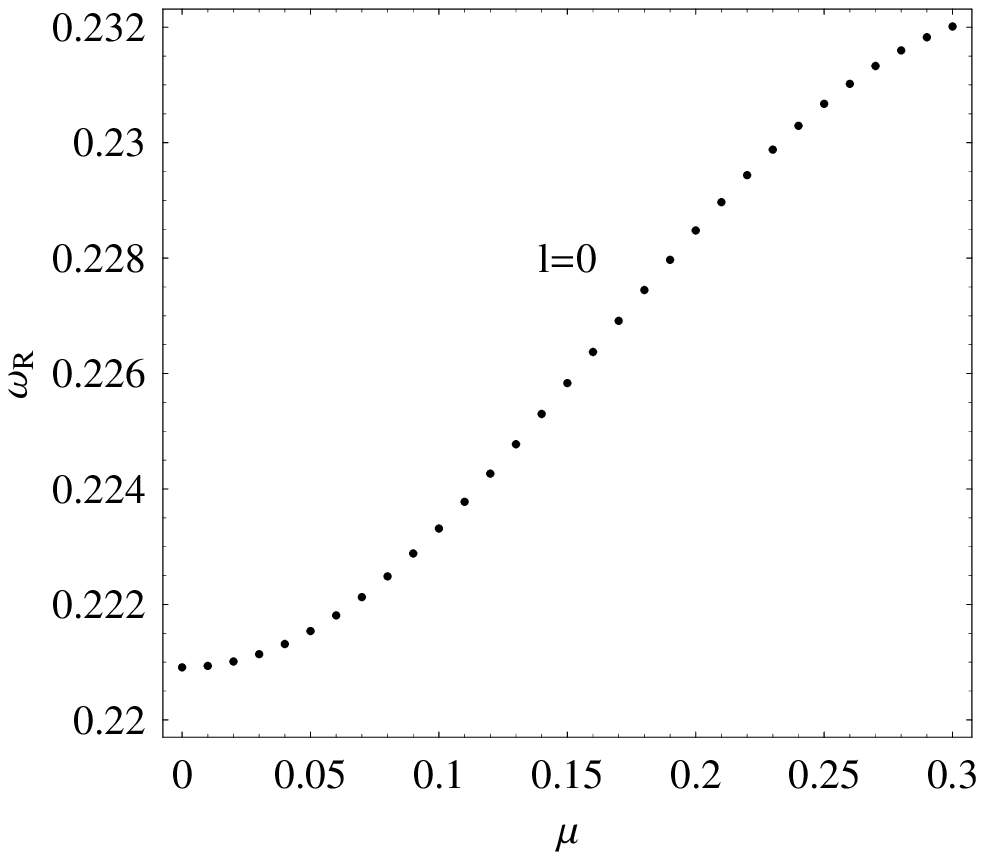}\ \ \ \ \ \ \includegraphics[width=6cm]{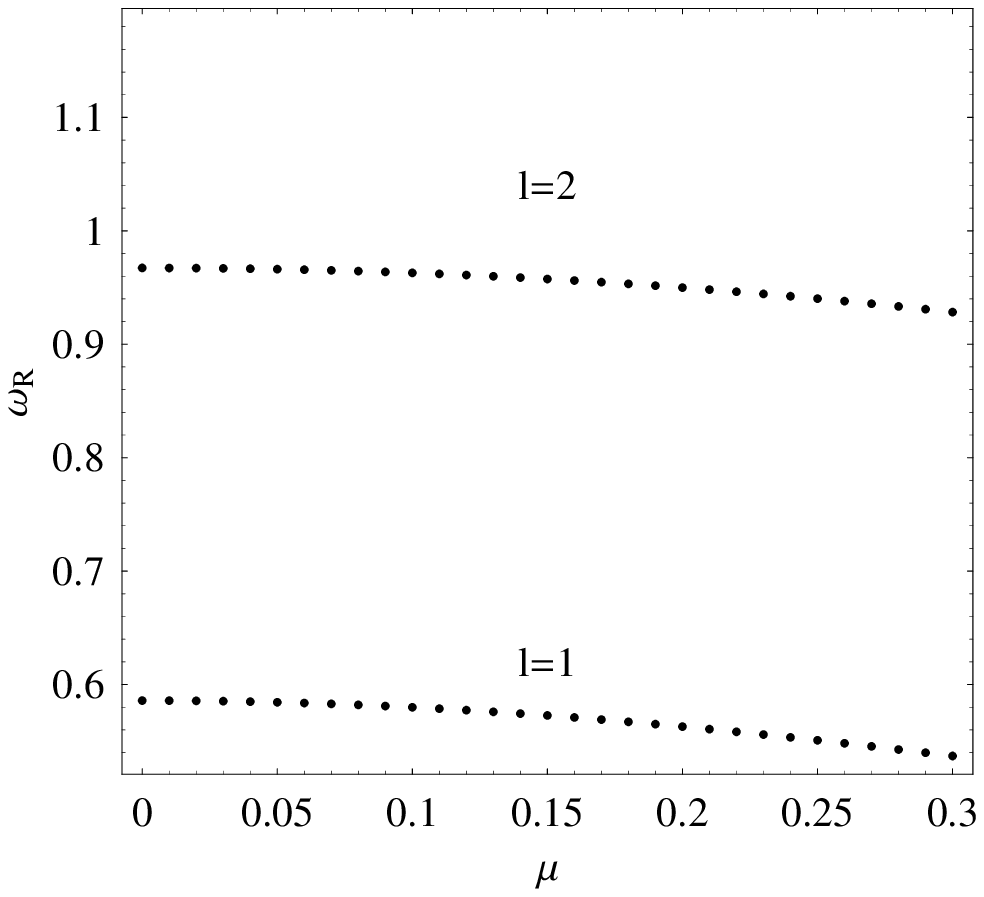}
\caption{Change of the real part quasinormal frequencies ($n=0$) for
the phantom scalar perturbations with the parameter $\mu$ for fixed
$l$.}
\end{center}
\end{figure}
\begin{figure}[ht]
\begin{center}
\includegraphics[width=8cm]{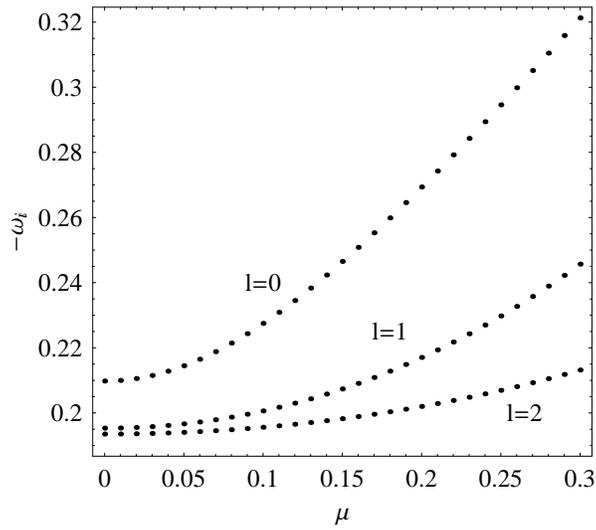}
\caption{Change of the imaginary part quasinormal frequencies
($n=0$) for the phantom scalar perturbations with the parameter
$\mu$ for fixed $l$.}
\end{center}
\end{figure}
In table I, we list the fundamental quasinormal frequencies of
phantom scalar perturbation field for fixed $l=0$, $l=1$ and $l=2$
in the Schwarzschild black hole spacetime. From the table I and
figures (1) and (2), we find that with the increase of the mass
$\mu$ the real parts increase for $l=0$ and decrease for $l=1 $ and
$l=2$. The absolute value of imaginary parts for all $l$ increase,
which can be explained by that the bigger $\mu$ leads to lower peak
of the potential and thus it is easier for the wave to be absorbed
into the black hole. The dependence of quasinormal modes on the mass
$\mu$ is different from that of the usual scalar field since their
effective potentials are quite a different in the black hole
spacetime.

Now we will extend to study the late-time behavior of phantom scalar
perturbations in the Schwarzschild black hole spacetime by using of
the time domain method \cite{T1}.

Adopting the null coordinates $u=t-r_*$ and $v=t+r_*$, the wave
equation
\begin{eqnarray}
-\frac{\partial^2\psi}{\partial t^2}+\frac{\partial^2\psi}{\partial
r_*^2}=V(r)\psi,
\end{eqnarray}
 can be recast as
 \begin{eqnarray}
4\frac{\partial^2\psi}{\partial u\partial
v}+V(r)\psi=0,\label{wbes1}
\end{eqnarray}
The two-dimensional wave equation (\ref{wbes1}) can be integrated
numerically, using for example the finite difference method
suggested in \cite{T1}. Using Taylor's theorem, it is discretized as
\begin{eqnarray}
\psi_N=\psi_E+\psi_W-\psi_S-\delta u\delta v
V(\frac{v_N+v_W-u_N-u_E}{4})\frac{\psi_W+\psi_E}{8}+O(\epsilon^4)=0,\label{wbes2}
\end{eqnarray}
where the points $N$, $S$, $E$ and $W$ from a null rectangle with
relative positions as: $N$: $(u+\delta u, v+\delta v)$, $W$: $(u +
\delta u, v)$, $E$: $(u, v + \delta v)$ and $S$: $(u, v)$. The
parameter $\epsilon$ is an overall grid scalar factor, so that
$\delta u\sim\delta v\sim\epsilon$. Considering that the behavior of
the wave function is not sensitive to the choice of initial data, we
set $\psi(u, v=v_0)=0$ and use a Gaussian pulse as an initial
perturbation, centered on $v_c$ and with width $\sigma$ on $u=u_0$
as
\begin{eqnarray}
\psi(u=u_0,v)=\text{exp}[-\frac{(v-v_c)^2}{2\sigma^2}].\label{gauss}
\end{eqnarray}
\begin{figure}[ht]
\begin{center}
\includegraphics[width=6cm]{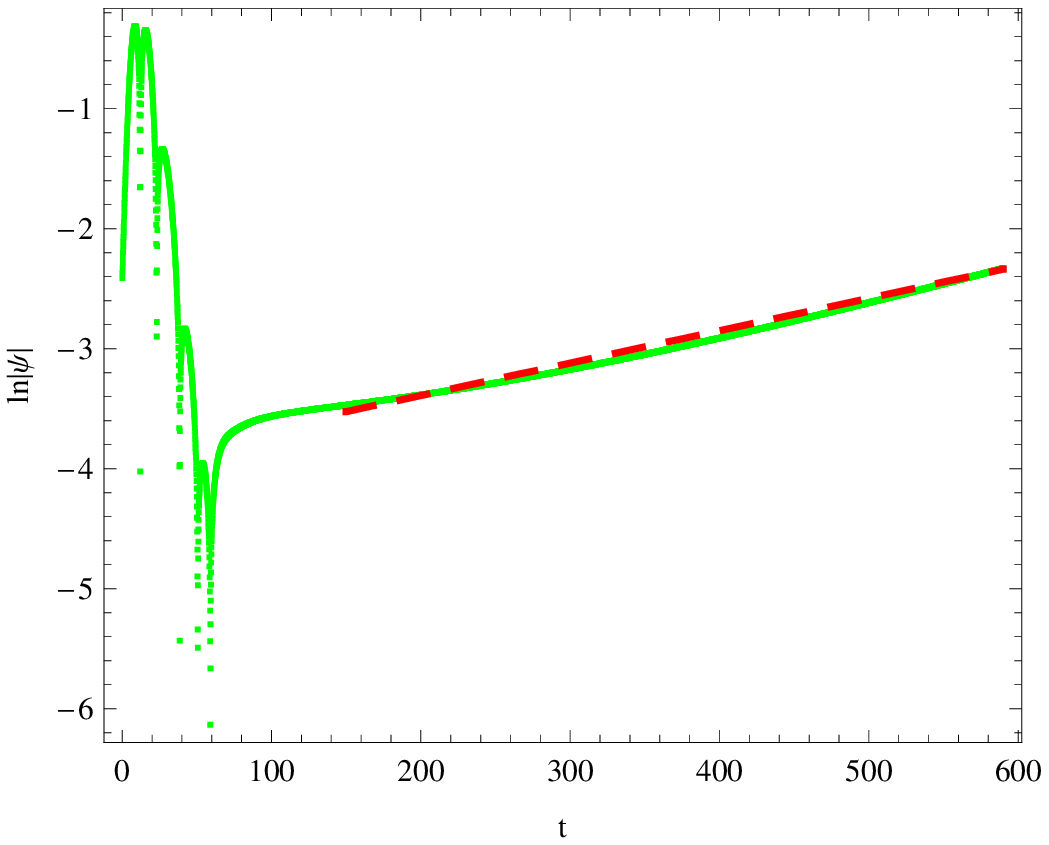}\;\;\;\;
\includegraphics[width=6cm]{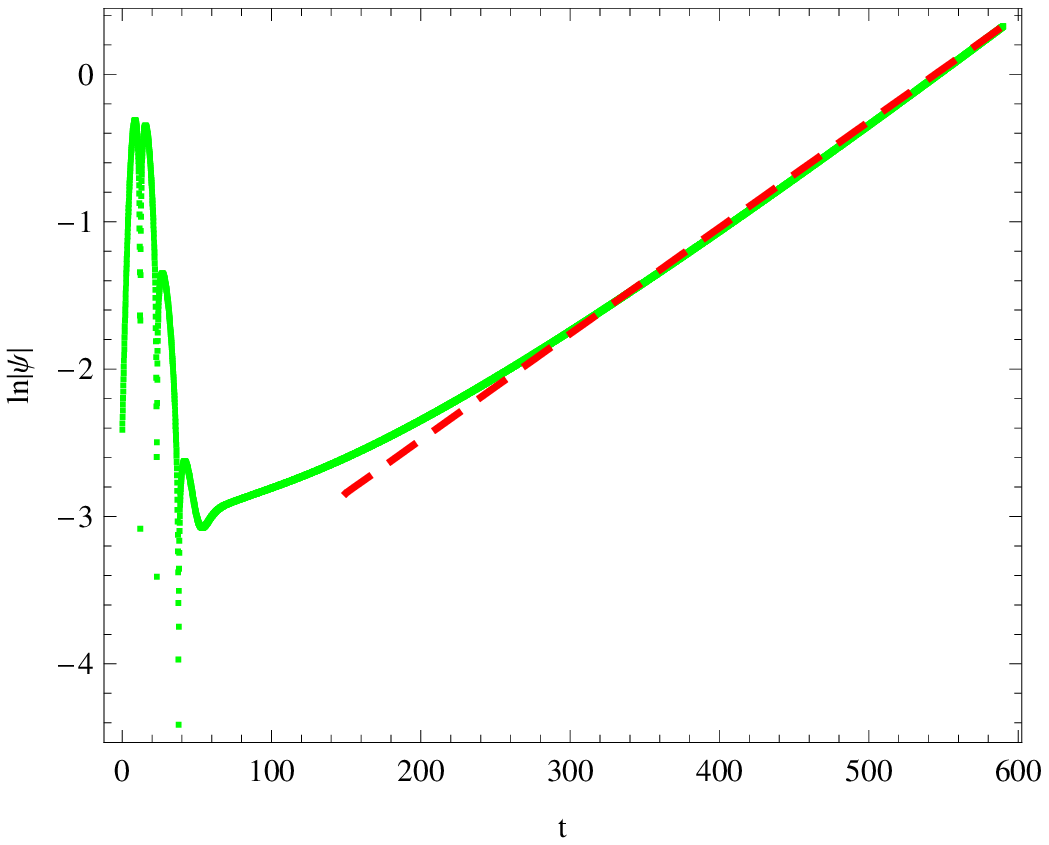} \caption{ The late-time
behaviors of the phantom scalar perturbations for fixed $l=0$, the
left and the right are for $\mu=0.01$ and $0.02$, respectively. The
dashed line denotes the function $\ln \psi\sim \alpha \mu
t-4l-\beta$. We set the numerical constants $\alpha=0.27$,
$\beta=3.93$ in the left figure and $\alpha=0.36$, $\beta=3.92$ in
the right one. The constants in the Gauss pulse (\ref{gauss})
$v_c=10$ and $\sigma=3$.}
\end{center}
\end{figure}
\begin{figure}[ht]
\begin{center}
\includegraphics[width=6cm]{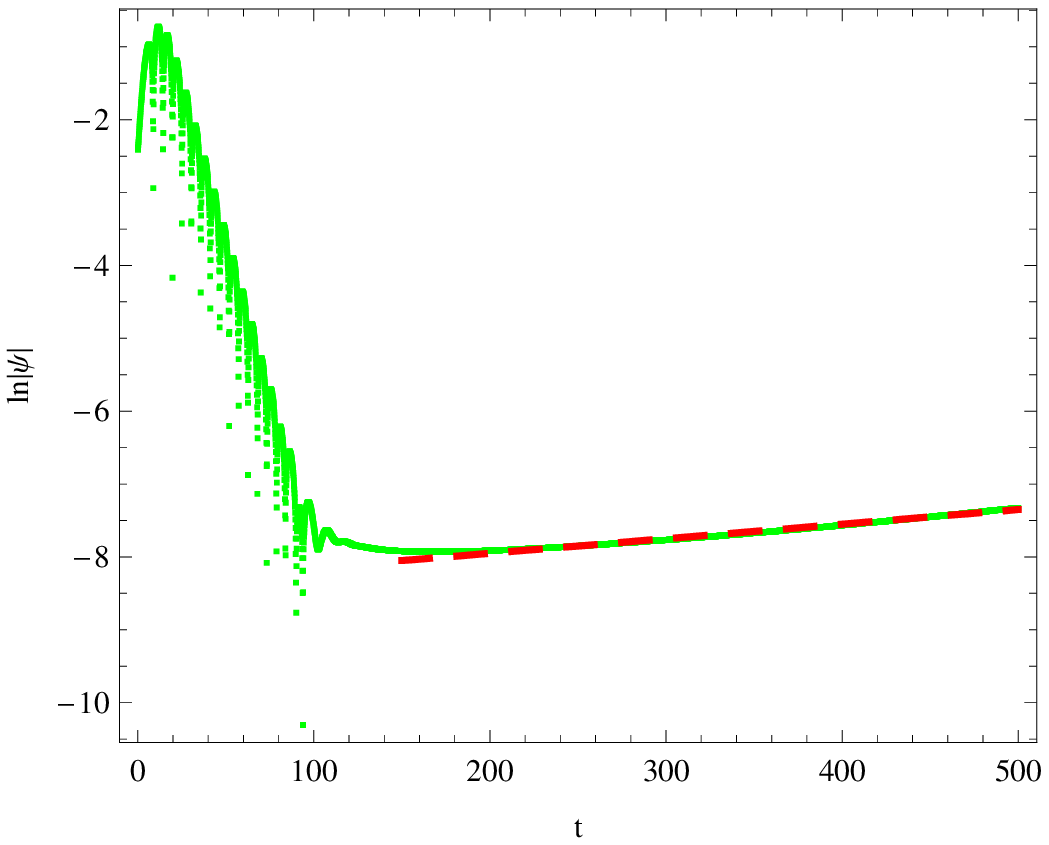}\;\;\;\;
\includegraphics[width=6cm]{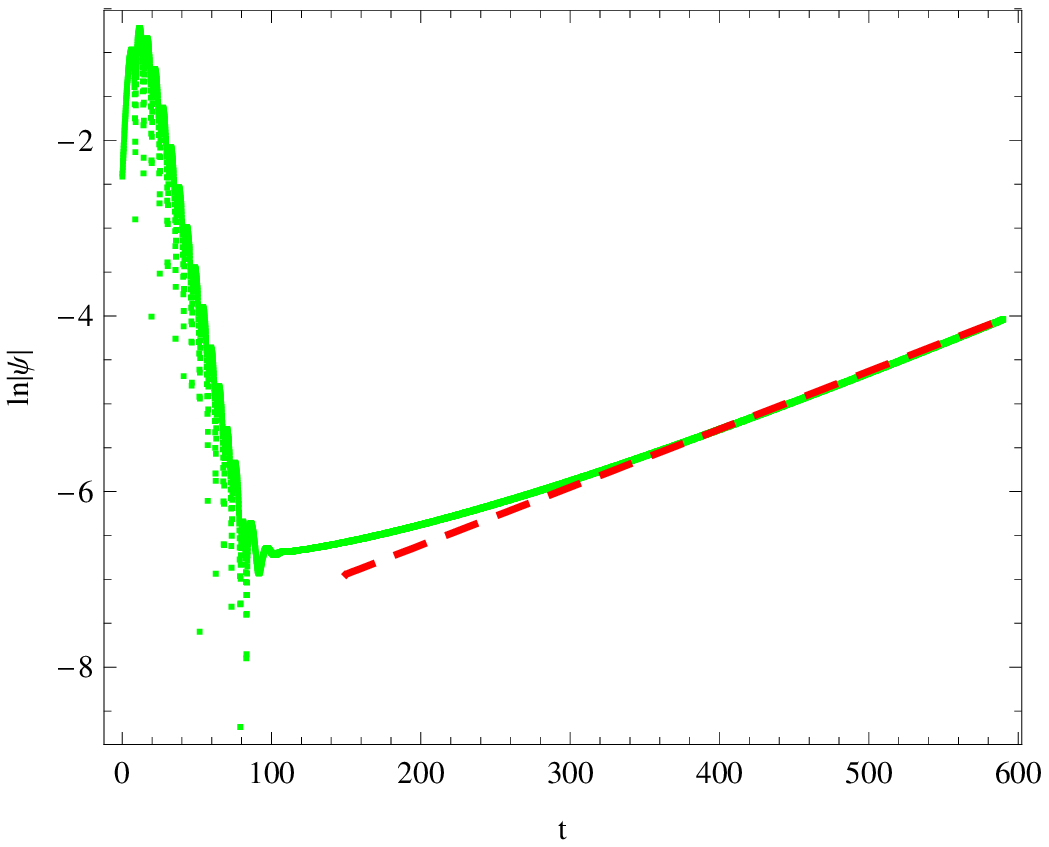} \caption{ The late-time
behaviors of the phantom scalar perturbations for fixed $l=1$, the
left and the right are for $\mu=0.01$ and $0.02$, respectively. The
dashed line denotes the function $\ln \psi\sim \alpha \mu
t-4l-\beta$. We set the numerical constants $\alpha=0.20$,
$\beta=4.32$ in the left figure and $\alpha=0.32$, $\beta=3.84$ in
the right one. The constants in the Gauss pulse (\ref{gauss})
$v_c=10$ and $\sigma=3$.}
\end{center}
\end{figure}
\begin{figure}[ht]
\begin{center}
\includegraphics[width=6cm]{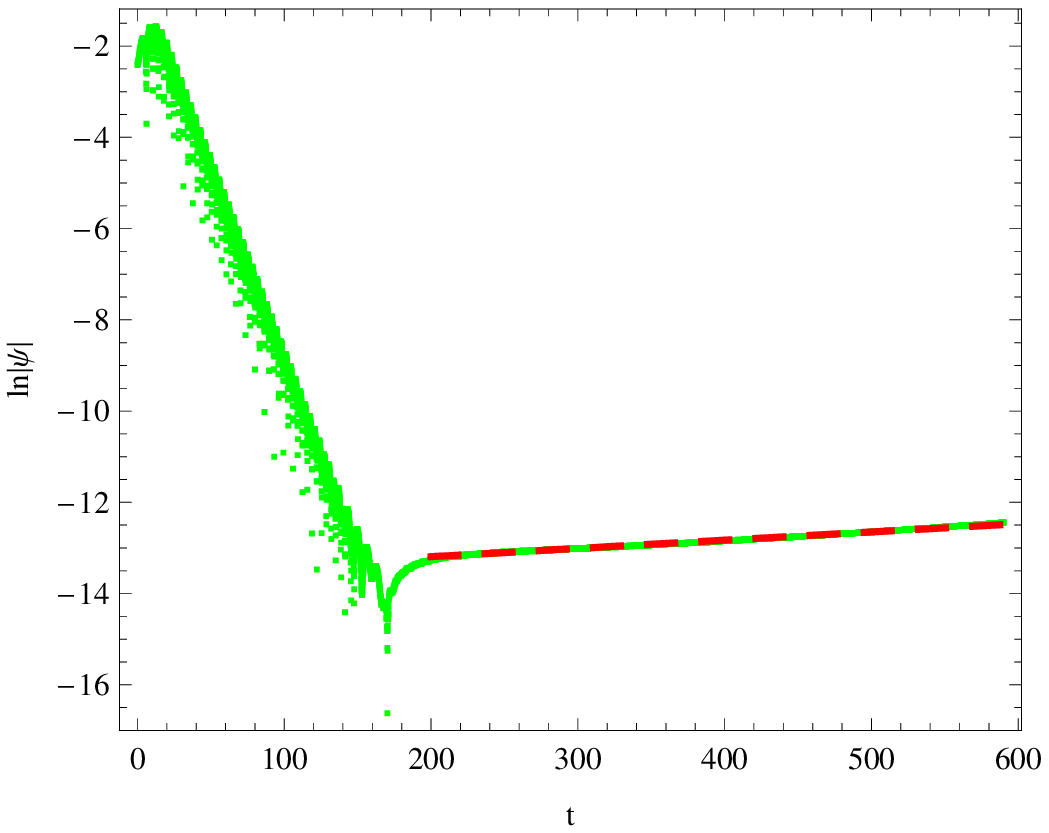}\;\;\;\;
\includegraphics[width=6cm]{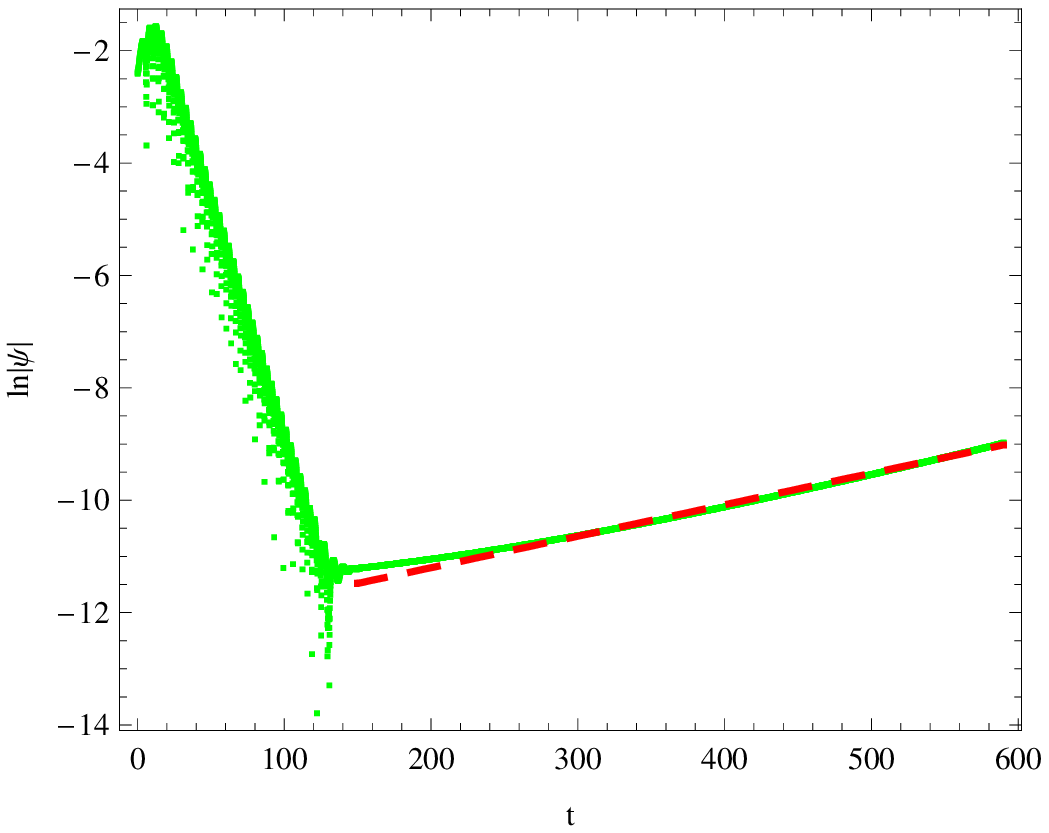} \caption{ The late-time
behaviors of the phantom scalar perturbations for fixed $l=2$, the
left and the right are for $\mu=0.01$ and $0.02$, respectively. The
dashed line denotes the function $\ln \psi\sim \alpha \mu
t-4l-\beta$. We set the numerical constants $\alpha=0.18$,
$\beta=5.55$ in the left figure and $\alpha=0.28$, $\beta=4.30$ in
the right one. The constants in the Gauss pulse (\ref{gauss})
$v_c=10$ and $\sigma=3$.}
\end{center}
\end{figure}
Here we confirm numerically the peculiar properties in late-time
evolution of phantom scalar perturbations in the Schwarzschild black
hole spacetime. In figures (4), (5) and $(6)$, we plot the
evolutions of the phantom scalar perturbations for fixed $l=0$,
$l=1$ and $l=2$, respectively. Unlike the usual massive scalar
perturbations, the phantom scalar field after undergoing the
quasinormal modes stage does not decay but grow with the time. The
asymptotic behaviors of wave function for the phantom scalar field
in the Schwarzschild black hole spacetime can be fitted best by
\begin{eqnarray}
\psi\sim e^{\alpha \mu t-4l-\beta},
\end{eqnarray}
where $\alpha$ and $\beta$ are two numerical constant. It means that
the phantom scalar perturbation grows with exponential rate in the
late-time evolution. This behavior can be attributed to that the
effective potential is negative at the spatial infinity shown in
fig.(1), which implies that the wave outside the black hole gains
energy from the spacetime. For the larger $\mu$, the scalar
perturbation grows more faster since the larger $\mu$ corresponds to
more negative potential. It is quite different from that of the
usual massive scalar perturbations in a black hole spacetime which
decay with the oscillatory inverse power-law behavior
$t^{-\gamma}\sin{\mu t}$ \cite{ml1,ml2}. Moreover, the growing modes
of the late-time tails caused by the negative effective potential
was also observed in \cite{bwr}.

The exponential growth of the phantom scalar perturbation in the
late-time evolution also means that it is unstable in the black hole
spacetime and its energy density will increase with the time, which
is similar to that of in the Einstein cosmology. Our result also
agrees with that obtained in the phantom energy accretion of black
hole \cite{Eph}, where the black hole mass will be decreased and the
energy density of the phantom will increase.

In summary we examine the wave dynamics of the phantom scalar
perturbation in the background of Schwarzschild black hole. Our
results show that due to the presence of the negative kinetic
energy, the properties of the wave dynamics of phantom scalar
perturbation are different from that of the usual massive scalar
field. In the late-time evolution the phantom field does not decay
but instead grows with an exponential rate.  It would be of interest
to generalize our study to other black hole spacetimes, such as
rotating black hole and stringy black holes etc. Work in this
direction will be reported in the future.

\begin{acknowledgments}

We are grateful to Professor Bin Wang for his help. This work was
partially supported by the National Natural Science Foundation of
China under Grant No.10875041; the Scientific Research Fund of Hunan
Provincial Education Department Grant No.07B043 and the construct
program of key disciplines in Hunan Province. J. L. Jing's work was
partially supported by the National Natural Science Foundation of
China under Grant No.10675045; the FANEDD under Grant No. 200317;
and the Hunan Provincial Natural Science Foundation of China under
Grant No.08JJ3010.
\end{acknowledgments}

\end{document}